\documentstyle[12pt]{article}
\def\be{\begin{equation}}
\def\ee{\end{equation}}
\begin{document}

\title{ON THE GRAVITOELECTROMAGNETIC STRESS-ENERGY TENSOR}

\author{Bahram Mashhoon and James C. McClune\\
								Department of Physics and Astronomy\\
        University of Missouri - Columbia\\
        Columbia, MO 65211, USA
                                            \\
 \and
                                            \\
        Hernando Quevedo\\
        Instituto de Ciencias Nucleares\\
        Universidad Nacional Aut\'onoma de M\'exico\\
        A.P. 70 - 543\\
        04510 M\'exico D. F., M\'exico }
\date{}
\baselineskip 18 pt

\maketitle
\begin {abstract} 
We study the pseudo-local gravitoelectromagnetic stress-energy tensor for an arbitrary gravitational field
within the framework of general relativity.  It is  shown that there exists a current of gravitational energy around a 
rotating mass.  This gravitational analog of the Poynting flux is evaluated for certain classes of observers in the Kerr
field.
\end {abstract}
\newpage

\maketitle

\section*{1 Introduction} Imagine a packet of long wavelength gravitational waves incident on a Keplerian binary system.  In
the lowest (quadrupole) approximation, the waves exchange energy and angular momentum with the self-gravitating system.  To
describe this interaction in detail, it is evident that an essentially local measure of the energy and stress carried by the
waves is necessary.  Therefore, there must be a  somewhat local way to describe the energy and momentum stored in the gravitational
field.  A partial (i.e.  gravitoelectromagnetic) solution of this general
problem within the framework of  general relativity is essentially contained in a recent paper \cite{1}.  The  present work is
concerned with the extension of our approach and some of its  general consequences.  Our paper on the gravitational
superenergy tensor 
\cite{1} should be consulted for much of the background material; however, we  have attempted to make the present paper
essentially self-contained.

The standard (Landau-Lifshitz) gravitational stress-energy {\it pseudotensor} is useful in general relativity only in a global
sense for asymptotically flat spacetimes. For many applications, however, a {\it local} measure of the  stress-energy content
of an arbitrary gravitational field would be helpful. It is possible to provide such a measure --- though only as an
approximation --- for the gravitoelectromagnetic part of the field as shown in our recent paper 
\cite{1}. As expected, the gravitoelectromagnetic stress-energy tensor has  general properties that are rather similar to
those of the Maxwell stress-energy tensor in electrodynamics. For a Ricci-flat spacetime, this  approach provides an average
measure of the gravitational stress-energy content that is proportional to the Bel-Robinson tensor. Thus our derivation of the 
gravitational 
stress-energy tensor provides a natural physical interpretation of the  Bel-Robinson tensor that has been used frequently in
numerical relativity \cite{2}.  

The curved spacetimes of general relativity are not asymptotically flat in  general; therefore, the usual concepts of energy,
momentum, and stress do not make sense in the standard interpretation of general relativity. Nevertheless, it is possible to
introduce a pseudo-local gravitoelectromagnetic (``GEM")  stress-energy tensor via a certain averaging procedure.  This
approach is briefly discussed in the next section and the general properties of the GEM tensor are described.  In sections 
3 and 4, special gravitational fields are considered --- such as the charged Kerr-Taub-NUT spacetime --- and their GEM stress-energy tensors are evaluated.  Moreover, the existence of a steady current of gravitational field energy around a rotating mass i

s predicted and some of its properties are briefly studied in section 5. We employ units such that the speed of light $c=1$; furthermore, the spacetime metric has signature +2, Greek indices run from 0 to 3 and Latin indices run from 1 to 3.

\section*{2 GEM Stress-Energy Tensor}

Consider a free test observer in a gravitational field following a worldline 
${\cal C}$ and let $\tau$ be the proper time along ${\cal C}$.  In a Fermi coordinate  system $X^\alpha = (\tau,{\bf X})$
along the geodesic path of the observer, the physical content of the spacetime interval can be described in terms of a 
gravitoelectric potential $\Phi$, a gravitomagnetic vector potential
${\bf A}$ and a spatial tensor potential $S$.  In fact, the metric in the Fermi frame is given by ${^F}\!g_{\mu\nu} =
\eta_{\mu\nu} + {^F}\!h_{\mu\nu}$, where
${^F}\!h_{00} =  2\Phi$, $^{F}\!h_{0i}=-2A_{i}$ and $^{F}\!h_{ij}=S_{ij}$.  To lowest order in spatial Fermi coordinates away from
${\cal C}$,
\be
\Phi=-\frac{1}{2} \:{}^F \!R_{0i0j}(\tau)X^{i}X^{j},
\ee
\be
A_{i}=\frac{1}{3} \:{}^{F} \!R_{0jik}(\tau)X^{j}X^{k},
\ee
\be 
{S}_{ij}=-\frac{1}{3} \:{}^{F}\!R_{ikjl}(\tau)X^{k}X^{l},
\ee  where $^{F} \!R_{\alpha\beta\gamma\delta}=R_{\mu\nu\rho\sigma}
\lambda^{\mu}_{(\alpha)}\lambda^{\nu}_{(\beta)}\lambda^{\rho}_{(\gamma)}\lambda^{\sigma}_{(\delta)}$ is the spacetime curvature
measured by the observer, i.e. it is the projection of the Riemann curvature tensor along the nonrotating tetrad system
$\lambda^{\mu}$$_{(\alpha)}(\tau)$ carried by the observer at $C^{\mu}:(\tau, {\bf 0})$ in the Fermi system.  In general, the
locally measured components of the Riemann curvature tensor, $R_{\mu\nu\rho\sigma}
\lambda^{\mu}_{(\alpha)}\lambda^{\nu}_{(\beta)}\lambda^{\rho}_{(\gamma)}
\lambda^{\sigma}_{(\delta)}$,  can be represented as a symmetric $ 6 \times 6$ matrix ${\cal R}=({\cal R}_{IJ})$, where the
indices $I$ and $J$ range over the  set
$(01, 02, 03, 23, 31, 12)$. The symmetries of the Riemann tensor make it possible to write
\be {\cal R} = \left[ \begin{array} {rr} E \ \   &  B \\ B^{T}  &  D \\
\end{array} \right],
\ee where $E$ and $D$ are symmetric $3 \times 3$ matrices and $B$ is traceless. Here $E$ and $B$ represent the
``electric" and ``magnetic" components of the spacetime curvature, respectively, and $D$ represents the ``spatial" components.
The curvature of the gravitational field is thus determined  by the three  matrices $E,\ B$ and $D$.  In a Ricci-flat
spacetime, $R_{\mu\nu}=0$, $D=-E$ so that the spatial components of the curvature are completely determined by the
electric components; moreover, $E$ is traceless and $B$ is symmetric $\left(B^{T}=B\right)$.

Using the potentials $\Phi$ and ${\bf A}$, it is natural to define the gravitoelectric and gravitomagnetic fields in complete
analogy with electrodynamics.  Keeping only the lowest order terms in the spatial Fermi coordinates, we find
\be {\cal E}_{i}=\:{}^{F}\!R_{0i0j}(\tau)X^{j},
\ee
\be {\cal B}_{i}=-\frac{1}{2}\epsilon_{ijk}\:{}^F\!R_{jk0l}(\tau)X^{l}.
\ee It turns out that these fields can be combined in a gravitoelectromagnetic ``Faraday'' tensor given by
\be {\cal F}_{\alpha\beta}=-\:{}^{F}\!R_{\alpha\beta0l}(\tau)X^{l}.
\ee It is then natural to construct from the GEM field tensor (7) the corresponding ``Maxwell'' stress-energy tensor
\be {\cal T}^{\alpha\beta}=\frac{1}{4\pi}\left({\cal F}^{\alpha}_{\ \gamma}{\cal
F}^{\beta\gamma}-\frac{1}{4}\eta^{\alpha\beta}{\cal F}_{\gamma\delta}{\cal F}^{\gamma\delta}\right)
\ee that is symmetric and traceless by construction just as in electromagnetism.  We can also define the GEM angular
momentum density tensor
\be
{\cal J}^{\alpha\beta\gamma}=\left(X^{\alpha}-C^{\alpha}\right) {\cal T}^{\beta\gamma}-\left(X^{\beta}-C^{\beta}\right) {\cal
T}^{\alpha\gamma}, \ee
where $C^{\alpha}:(\tau, {\bf 0})$ denote the coordinates of the fiducial test observer at the center of the
Fermi system (i.e. the spatial origin of Fermi coordinates).  Thus ${\cal J}^{\alpha\beta\gamma}$ vanishes at the spacetime
position of the test observer by construction.  The GEM angular momentum tensor is then defined by 
\be
J^{\alpha\beta}=\int {\cal J}^{\alpha\beta\gamma} d^{3} \Sigma_{\gamma}
\ee as in standard field theory, except that the relevant 3D volume must be within the Fermi system and
consistent with the approximation scheme under consideration here.

The stress-energy tensor (8) has the property that it vanishes along the worldline ${\cal C}$ by Einstein's principle of
equivalence; however, it is in general nonzero on geodesic paths in the neighborhood of the fiducial worldline. Measurement of
such field quantities does not occur at a point  in space; in fact, an averaging process is indispensable. An interesting and
useful measure of the gravitoelectromagnetic stress-energy in the immediate vicinity of the observer may be obtained by
averaging the  stress-energy tensor over a sphere of radius
$\epsilon L$ at each instant  of proper time along the fiducial path. Here $\epsilon$, $0<\epsilon \ll 1$, is a small parameter
and $L$ is an intrinsic length--scale characteristic  of the gravitational field.  The form of the tensor which is averaged in the
Fermi frame  is such that the same result is in effect obtained whether one averages over the surface or the volume of an
infinitesimal sphere about the spatial origin of Fermi coordinates, since the difference could be absorbed in the definition
of $L$.  The resulting average stress-energy tensor is  then physically defined only up to an overall positive  constant scale factor.

It follows from these considerations that the average gravitoelectromagnetic stress-energy content of the field as determined by the observer at the event under
consideration is given by the projection of a symmetric traceless stress-energy tensor
$\tilde T_{\mu\nu}$ on the observer's frame, i.e.
\begin{eqnarray}
\tilde{T}_{(\alpha)(\beta)} = {L^{2} \over G_{0}} 
\tilde{T}_{\mu\nu\rho\sigma}\lambda^{\mu}_{(\alpha)}\lambda^{\nu}_{(\beta)}
\lambda^{\rho}_{(0)}\lambda^{\sigma}_{(0)},
\end{eqnarray}
where $L$ is the constant invariant length characteristic of the gravitational field, $G_0$ is Newton's constant and
$\tilde{T}_{\mu\nu\rho\sigma}$ is given by
\be
        \tilde{T}_{\mu\nu\rho\sigma}={1\over 2}     \left( R_{\mu\xi\rho\zeta}R_{\nu\ \sigma}^{\ \xi\ \zeta}+
R_{\mu\xi\sigma\zeta}R_{\nu\
\rho}^{\ \xi\ \zeta}   \right) -{1\over 4}g_{\mu\nu}R_{\alpha\beta\rho\gamma} R^{\alpha\beta\ \,\gamma}_{\ \ \ \sigma\ } .
\ee
This tensor is symmetric and traceless in its first pair of indices and symmetric in its second pair of indices and
reduces to the completely symmetric and traceless Bel-Robinson tensor, $T_{\mu\nu\rho\sigma}$, for a Ricci-flat spacetime (cf.
appendix A).  Then $\tilde{T}_{(\alpha)(\beta)}$ reduces in turn to ${T}_{(\alpha)(\beta)}$, which we call the ``gravitational
stress-energy tensor" since the curvature is completely characterized in this case by its GEM part.

Equations (11) and (12) originate from local dynamical considerations [1]. In a Fermi frame established along the worldline of a
geodesic observer, the measured gravitoelectric and gravitomagnetic fields to linear order in the spatial distance away from 
the fiducial path can be combined to form a gravitoelectromagnetic Faraday tensor and the corresponding GEM Maxwell stresses
and angular momentum densities.  
If (9) is integrated over the volume of the infinitesimal sphere, i.e. if in equation (10) we integrate at a
given $\tau$ over the sphere of center $C^{\mu}:(\tau, {\bf 0})$ and radius $\epsilon$, then we find that
$J^{\alpha\beta}=0$.  This means that the local center of mass of the sphere is indeed $C^{\mu}:(\tau, {\bf 0})$ as expected;
moreover, the field angular momentum
$J^{ij}$ contained in this sphere vanishes at the linear order of approximation for the fields (5) and (6) under consideration here.

We find that in general
\begin{eqnarray}
\tilde{T}_{(0)(0)} = {L^{2} \over 2G_{0}}\ {\rm tr}(E^{2}+B^{T}B), 
\end{eqnarray}
\begin{eqnarray}
\tilde{T}_{(0)(i)}={L^{2} \over G_{0}}
\ \epsilon_{ijk}(EB)_{jk},
\end{eqnarray}
\begin{eqnarray}
\tilde{T}_{(i)(j)} = {L^{2} \over 2G_{0}} [-2 (E^{2}+B^{T}B)_{ij} +
\delta_{ij}\ {\rm tr}(E^{2}+B^{T}B)] ,
\end{eqnarray} so that $\tilde{T}_{(\alpha)(\beta)}$ only contains the ``electric" and ``magnetic" parts of the spacetime
curvature and not the ``spatial" part. It is immediately clear from equation (13) that $\tilde T_{(0)(0)}$ is positive
definite; hence, $\tilde T_{(0)(0)}=0$ implies that
$E=0$ and $B=0$  and the spacetime curvature is thus purely ``spatial" for the observer in this case. Moreover, by writing the matrices $E$ and
$B$ in equations (13) - (15) in component form and repeated application of the simple relations
$\vert \eta+\nu\vert\le\vert\eta\vert + \vert\nu\vert$ and 
$\vert\eta\nu\vert\le {1\over 2}(\eta^{2}+\nu^{2})$ for arbitrary real numbers 
$\eta$ and $\nu$, it is straightforward to show that
\begin{eqnarray}
\vert\tilde{T}_{(0)(i)}\vert \le \tilde{T}_{(0)(0)},
\end{eqnarray}
and
\begin{eqnarray}
\vert\tilde{T}_{(i)(j)}\vert\le\tilde{T}_{(0)(0)}.
\end{eqnarray} Thus the gravitational Poynting vector is always timelike or null and the absolute magnitude of a
gravitoelectromagnetic stress is always bounded above by the local average density. These relations are reminiscent of
Maxwell's electrodynamics.

It is interesting to note that in general
\begin{eqnarray} {1 \over 4}R_{\mu\nu\rho\sigma}R^{\mu\nu\rho\sigma} = {\rm tr} (E^{2}-2B^{T}B+D^{2}), 
\end{eqnarray}
and
\begin{eqnarray} {1 \over 16}R_{\alpha\beta}^{\ \ \
\mu\nu}R^{\alpha\beta\rho\sigma} e_{\mu\nu\rho\sigma} = {\rm tr}(EB-BD),  
\end{eqnarray}
where $e_{\mu\nu\rho\sigma}$ is the alternating tensor
$e_{\mu\nu\rho\sigma} = (-g)^{1/2} \epsilon_{\mu\nu\rho\sigma}$ with
$\epsilon_{0123}\equiv 1$.

Let us now assume that $R_{\mu\nu}=0$. In this case, $D=-E$,  $E$ is traceless and $B$ is symmetric. Thus $E$ and $B$
characterize the whole gravitational field in this case. This is illustrated in the next section via approximate 
gravitational field solutions that are Ricci flat. Moreover, in the
Ricci-flat case, as noted already by Matte \cite{4}, the curvature invariants (18) and (19) divided by 2  take
the forms tr$(E^{2}-B^{2})$ and
tr$(EB)$, respectively, that are familiar from electrodynamics.

The gravitoelectromagnetic stress-energy tensor $\tilde{T}_{\mu\nu}$ reduces in the Ricci-flat case to the gravitational
stress-energy tensor $T_{\mu\nu}$ discussed previously \cite{1}. The Riemann tensor reduces to the Weyl tensor in the latter
case; therefore, the connection between $\tilde{T}_{\mu\nu}$ and
$T_{\mu\nu}$ can be worked out in general using
\be
     R_{\mu\nu\rho\sigma}=C_{\mu\nu\rho\sigma}+ 
        {1\over 2}       \left( 
          R_{\mu\rho}g_{\nu\sigma}+ R_{\nu\sigma}g_{\mu\rho}
         -R_{\mu\sigma}g_{\nu\rho}- R_{\nu\rho}g_{\mu\sigma}    \right)
         -{1\over 6} R   \left( 
         g_{\mu\rho}g_{\nu\sigma}-g_{\mu\sigma}g_{\nu\rho}      \right) ,
\ee where $R_{\mu\nu}$ and $R$ are given in terms of the stress-energy tensor of the source of the field via the
gravitational field equations. It is more interesting, however, to work out $T_{\mu\nu}$ explicitly for certain approximate
solutions of the gravitational field equations in which nonlinearities are neglected for the sake of simplicity.  This is
done in the next section.

\section*{3 Linear Gravitational Fields}

Let us first imagine linear gravitational waves given by $g_{\mu\nu}=
\eta_{\mu\nu}+h_{\mu\nu}$, where a gauge is chosen such that $h_{0\mu} = 0$,\, $ h^{ij}_{\ ,j}=0$ and tr$(h_{ij})=0$. All
static observers follow  geodesics of this gravitational field. The gravitational wave amplitudes
$h_{ij}(t,{\bf x})$ satisfy the wave equation; therefore, we consider for the sake of simplicity a plane monochromatic wave of
frequency $\omega_{g}$ propagating along the $x$-direction. Then, the gravitoelectric and  gravitomagnetic fields as measured
by geodesic observers at fixed spatial positions can be expressed in this case as
\be
         E = {1\over 2 }\omega^{2}_{g}  
\left[ \begin{array} {ccc}
                            0 & 0          & 0 \\
                            0 & h_{+}      & h_{\times} \\
                            0 & h_{\times} & -h_{+}
        \end{array} \right] ,
\qquad 
         B = {1\over 2 }\omega^{2}_{g}  
\left[ \begin{array} {ccc}
                            0 & 0          & 0 \\
                            0 & h_{\times} & -h_{+} \\
                            0 & -h_{+}      & -h_{\times}
        \end{array} \right] ,
\ee where $2E_{ij}=\omega^{2}_{g}\ h_{ij}$, so that $h_{+}$ and $h_{\times}$  represent the two independent
linear polarization states of the wave. It is clear that a wave packet may be formed by a simple superposition of the fields
given by equation (21). It follows from equations  (13)--(15) that for each monochromatic component
\be
      \left( T^{{\scriptscriptstyle}(\alpha)(\beta)} \right)  =
        {{L^2 \omega^4_g}\over {2G_0}}(h_{+}^2 +h_{\times}^2)
\left[ \begin{array}{llll}
                  1 & 1 & 0 & 0 \\
                  1 & 1 & 0 & 0 \\
                  0 & 0 & 0 & 0 \\ 
                  0 & 0 & 0 & 0 \end{array}  \right] ,
\ee which should be compared and contrasted with the corresponding result  obtained using the Landau-Lifshitz pseudotensor
$t^{L-L}_{\mu\nu}$ (cf. the appendix of \cite{5}). In fact, $t^{L-L}_{\mu\nu}$ is in general gauge dependent and its trace,
\begin{eqnarray}
  {\rm tr}(t^{L-L}_{\mu\nu})={1 \over 32\pi G_{0}} h_{\alpha\beta ,\gamma} h^{\alpha\beta ,\gamma},
\end{eqnarray} is nonzero \cite{5}. That is, equation (23) is nonzero for a general wave packet, but
vanishes for a plane wave. On the other hand, the local gravitational stress-energy tensor is gauge invariant and traceless
just as in Maxwell's electrodynamics.

Let us next imagine the post-Newtonian field of a system with mass $M$ and angular momentum $J$ given in the linear
approximation by the standard metric 
\begin{eqnarray} -ds^{2}=-\left(1-2{G_{0}M \over \vert {\bf x} \vert}\right) dt^{2}-4{G_{0}\ dt \over \vert {\bf x} \vert^{3}} 
\epsilon_{ijk}J^{i}x^{j}dx^{k} +\left(1+2{G_{0}M\over \vert {\bf x}
\vert}\right)\delta_{ij}dx^{i}dx^{j}.
\end{eqnarray} An observer that is initially at rest in this field will not remain at a constant position ${\bf x}$; however, this motion can be neglected in the linear order of approximation under discussion in this section. It follows that \cite{6}
\begin{eqnarray} E_{ij}={G_{0}M \over \vert {\bf x} \vert^{3}}
\left( \delta_{ij}-3\hat{x}^{i}\hat{x}^{j}\right),
\end{eqnarray} 
\begin{eqnarray} B_{ij}=-3{G_{0}J\over \vert {\bf x} \vert^{4}}
\left[\hat{x}^{i}\hat{J}^{j} +
\hat{x}^{j}\hat{J}^{i}+(\delta_{ij} -5\hat{x}^{i}\hat{x}^{j})\hat{{\bf x}}\cdot
\hat{{\bf J}}\right],
\end{eqnarray} where $\hat {\bf x} = {\bf x}/\vert {\bf x} \vert$  is the unit position vector of the static
observer under consideration here. We find that
\begin{eqnarray} T^{(0)(0)}={3G_{0}L^{2} \over \vert {\bf x} \vert^{6}}
\bigg\{M^{2}+3{J^{2}
\over \vert {\bf x} \vert^{2}} \left[1+2(\hat{{\bf x}}\cdot\hat{{\bf  J}})^{2}\right]\bigg\},
\end{eqnarray}
\begin{eqnarray} T^{(0)(i)}={9G_{0}L^{2}MJ \over \vert {\bf x} \vert^{7}}
\left(\hat{{\bf J}} \times \hat{{\bf x}}\right)^{i},
\end{eqnarray}
\begin{eqnarray} T^{(i)(j)} & = & {3G_{0}L^{2}M^2 \over \vert {\bf x}
\vert^{6}}   
       \left({2\over 3}\delta_{ij}  -  \hat{x}^{i}\hat{x}^{j} \right) 
    +  {  9G_0L^2J^2 \over \vert {\bf x} \vert^8  }
\big[
  \delta_{ij} -\hat{x}^{i}\hat{x}^{j}- \hat{J}^{i}\hat{J}^{j} 
    \nonumber 
              \\ 
      \qquad \qquad &+&
  2(\hat{{\bf x}}\cdot \hat{{\bf J}})(\hat{x}^{i}\hat{J}^{j}  +
\hat{x}^{j}\hat{J}^{i}) +(\hat{{\bf x}}\cdot\hat{{\bf J}})^{2}(\delta_{ij} -5\hat{x}^{i}\hat{x}^{j}) 
\big]  .
\end{eqnarray}

The gravitational Poynting vector (28) is in this case analogous to the familiar circumstance in electrodynamics
involving the exterior field of a static nearly spherical system with a net electric charge and a constant magnetic dipole
moment.  The analogous gravitational properties would be mass $M$ and spin $J$, respectively. The Poynting vector, which is the
cross product of a radial electric field and a dipolar magnetic field, indicates a steady energy flux around the source just as
equation (28) indicates a steady gravitational energy current around a mass flowing in  the same sense as the rotation
of the body. Let us briefly digress here and mention that in the electromagnetic case,  the Poynting energy flux produces a
gravitoelectromagnetic field even when the material source itself does not rotate
\cite{7}. That is, the gravitational field is caused by the total stress-energy tensor, which in this case would originate
from the static electromagnetic source  together with the electromagnetic field that involves the steady energy flux.

The local average stress-energy tensor (27)-(29)  should be useful in the post-Newtonian investigation of the
dynamics of particles in the exterior gravitational field of a rotating mass.

\section*{4 Kerr Field}

To investigate further the nature of the steady gravitational energy flux around a rotating mass, it proves interesting to
study the GEM stress-energy tensor for the Kerr field.  Indeed, the study of the generalized Kerr spacetime is important due
to its possible physical significance in connection with the complete gravitational collapse of matter \cite{8}; 
however, we limit our treatment here to the charged Kerr-Taub-NUT spacetime for the sake of simplicity.  To this
end, let us imagine a set of observers following geodesic paths with nonrotating tetrads along their paths.  Suppose that at
some initial instant of time, the tetrad frames coincide with the natural tetrad system $\lambda^{\mu}$$_{(\alpha)}$ of the
charged Kerr-Taub-NUT spacetime.  We are interested in the measured GEM stress-energy tensor at this initial time.  It turns
out that in terms of Schwarzschild-like coordinates $(t, r, \theta, \phi)$,
$\omega^{(\alpha)}=\lambda_\mu^{\:(\alpha)}dx^{\mu}$ are given by
\be
\omega^{(0)}=\left( \frac {\Delta}{\Sigma}\right)^\frac {1}{2}\:[dt-(a \:{\sin}^{2}\theta-2l {\cos}\,\theta)d\phi],
\ee
\be
\omega^{(1)}=\left( \frac {\Sigma}{\Delta} \right)^\frac {1}{2} dr,
\ee
\be
\omega^{(2)}=\Sigma^\frac{1}{2}d\theta,
\ee
\be
\omega^{(3)}=\Sigma^{- \frac{1}{2}} [(r^{2}+a^{2}+l^{2})d\phi - a\,dt]\:{\sin}\, \theta,
\ee
\be
\Delta=r^{2}+a^{2}-l^{2}-2mr+q^{2},
\ee
\be
\Sigma=r^{2}+(a \:{\cos}\,\theta+l)^{2}.
\ee
Here $m, a, l,$ and $q$ are, respectively, the mass, angular momentum per unit mass, Taub-NUT parameter and the charge of the
source.  Only {\it positive} square roots are intended throughout this paper.

The components of the GEM stress-energy tensor in this tetrad system can be determined from the components of the curvature
tensor.  It is convenient to express the latter in the ${SO}(3,C)$ representation given by $M+P-i(N+Q)$, where $M, N,
P$, and $Q$ are $3 \times 3$ matrices in the bivector representation of the curvature tensor (4):
\be
{\cal R} = \left[ \begin{array} {rr} M  &  N \\ N  & -M  \\ \end{array} \right] 
+ \left[ \begin{array} {rr} P   &  Q \\ Q  &  -P \\ \end{array} \right] .
\ee
For the tetrad given above, we find \cite{9} 
\be
M-iN=\left[\frac{m+il}{(r+i\chi)^{3}}-\frac{q^{2}}{\Sigma(r+i\chi)^{2}} \right] {\rm diag} (-2, 1, 1),
\ee
\be
P-iQ=\frac {q^{2}}{\Sigma^{2}}\:{\rm diag} (1, 0, 0),
\ee
where $\chi=a\:{\cos}\,\theta+l$.  The resulting GEM stress-energy tensor is then given by
$$
\frac{G_0}{L^2} \left( \tilde{T}^{(\mu)(\nu)} \right) = \frac{m^2+l^2}{\Sigma^3}\:{\rm diag}(3,-1,2,2) + \frac{q^2}{\Sigma^4}
\:{\rm diag}(2\alpha,2\beta,\alpha-\beta,\alpha-\beta) 
$$
\be
+ \frac{\gamma q^2}{\Sigma^5}\:{\rm diag}(1,-1,1,1),
\ee
where
\be
\alpha=\frac{3}{4}q^{2}-2l\chi,
\ee
\be
\beta=\frac{1}{4}q^{2}-2mr,
\ee
\be
\gamma=4r^{2}(q^{2}-2l\chi-2mr).
\ee
The diagonal form of the stress-energy tensor turns out to be consistent with our approximate treatment in section 3; in
fact, this point will be discussed further in section 5.

Let us next consider the pure Kerr geometry $(l=0, q=0)$ and evaluate the gravitational stress-energy tensor for specific test
observers.  To this end, we first consider a free observer moving along the axis of axial symmetry such that far from the
source $(r \rightarrow \infty)$ the observer has speed $\beta_{0}$.  The equations of motion of the observer are given by
\be
\frac{dt}{d\tau}=\gamma_{0}\frac{\Lambda^{2}}{\Delta},
\ee
\be
\frac{dr}{d\tau}=\pm\left(\gamma_{0}^{2}-1+\frac {2mr}{\Lambda^{2}} \right) ^{1/2},
\ee
where $\gamma_{0}$ is the Lorentz factor at infinity $\gamma_{0}=(1-\beta_{0}^{2})^{-1/2}$, and
\be
\Lambda=(r^{2} + a^{2})^{1/2}\; .
\ee
The spatial triad is so chosen that $\lambda^{\mu}_{(1)}$ is along the radial direction (i.e. the $z$-axis) and is given by
\be
\lambda^{\mu}_{(1)}=\left[\dot{r}\frac{\Lambda^{2}}\Delta, \gamma_{0}, 0, 0 \right],
\ee
where $\dot{r}=dr/d\tau$.  Due to rotational symmetry about the direction of motion, there is a simple degeneracy in the choice
of $\lambda^{\mu}_{(2)}$ and $\lambda^{\mu}_{(3)}$, which are independent directions on a sphere of radius $\Lambda$.  It
suffices that $\lambda^{\mu}_{(\alpha)}$ be an orthonormal tetrad system, then the curvature as measured by the observer is
given by
\be
E=\frac{mr(r^{2}-3a^{2})}{\Lambda^{6}}\:{\rm diag}(-2,1,1),\quad B=\frac{ma(3r^{2}-a^{2})}{\Lambda^{6}}\:{\rm diag}(-2,1,1),
\ee
so that the gravitoelectric and the gravitomagnetic parts of the curvature are {\it parallel}.  The gravitational 
stress-energy tensor in this case turns out to be
\be
({T}_{(\alpha)(\beta)})=\frac{L^{2}m^{2}}{G_{0}\Lambda^{6}}\:{\rm diag}(3,-1,2,2)
\ee
just as in (39) along the axis of symmetry for $l=0$ and $q=0$.  This diagonal tensor is always regular since the observer can
simply pass through the ring singularity.

It is important to recognize that equations (47)-(48) do not depend on $\beta_{0}$ at all.  This important circumstance is a
consequence of the fact that the axis of symmetry provides two {\it special tidal directions} of the Kerr field so that the
curvature is independent of any Lorentz boosts along the ingoing and outgoing directions [6].  It follows that 
$T_{(\alpha)(\beta)}$ would then be independent of $\beta_{0}$ as well.  This situation has a direct analog in
electrodynamics.  That is, for an electromagnetic field that is not null one can always find a Lorentz frame in which the
electric and magnetic fields are parallel.  Any boost along the common direction of the fields leaves the fields as well as
the corresponding electromagnetic stress-energy tensor invariant.  It follows that (48) is an example of a general result: 
there exist special tidal directions at each event in a vacuum spacetime of type $D$ in the Petrov classification such that
the curvature as well as the GEM stress-energy tensor remain invariant under boosts along these directions \cite{10,11}.

It is interesting to consider a free test observer falling from rest at infinity in the equatorial plane of a Kerr system
$(\beta_{0}=0, \theta=\pi / 2)$ with zero orbital angular momentum (``radial motion'').  The geodesic path of the observer
is given in $(t, r, \theta, \phi)$ coordinates by
\be
\dot{t}=\frac{1}{\Delta}\left(\Lambda^{2}+\frac{2ma^{2}}{r} \right),
\ee
\be
\dot{r}^{2}=\frac{2m\Lambda^{2}}{r^{3}},
\ee
\be
\dot{\phi}=\frac{2ma}{r\Delta} .
\ee
Both incoming as well as outgoing geodesics are described by these equations.  We consider a nonrotating spatial triad along
the path given by
\be
\lambda^{\mu}_{(1)}=\left[ \frac{\Lambda}{\Delta}r\dot{r}, \frac{\Lambda}{r}, 0, \frac {ar\dot{r}}{\Delta\Lambda}
\right],
\ee
\be
\lambda^{\mu}_{(2)}=\left[ 0, 0, \frac{1}{r}, 0 \right],
\ee
\be
\lambda^{\mu}_{(3)}=\left[ - \frac{2ma\Lambda}{r\Delta}, - \frac{a\dot{r}}{\Lambda}, 0, \frac{\Lambda}{\Delta}
\left(1-\frac{2m}{r} \right) \right],
\ee
such that as $r \rightarrow \infty$ they correspond to the spherical coordinate axes \cite{12}.  The Riemann tensor as measured by the
(ingoing or outgoing) observer is given by
\be
E=\frac{m}{r^3} {\rm diag}(-2,1,1) + \frac{3ma^2}{r^5} {\rm diag}(-1,1,0), \quad B=\frac{3ma\Lambda}{r^5}
\left[
\begin{array}{ccc} 0 & 1 & 0 \\ 
                   1 & 0 & 0 \\
                   0 & 0 & 0 \\  \end{array} \right] .
\ee
Similarly, the nonzero components of the gravitational stress-energy tensor are independent of the direction of motion of the observer
and are given by
\be
{T}^{(0)(0)}=\frac{3L^{2}m^{2}}{G_{0}r^{10}}\left( r^{4}+6r^{2}a^{2}+6a^{4} \right),
\ee
\be
{T}^{(1)(1)}=-\frac{L^{2}m^{2}}{G_{0}r^{8}} \left(r^{2}+3a^{2} \right),
\ee
\be
{T}^{(2)(2)}=\frac{L^{2}m^{2}}{G_{0}r^{8}} \left(2r^{2}+3a^{2} \right),
\ee
\be
{T}^{(3)(3)}=\frac{2L^{2}m^{2}}{G_{0}r^{10}} \left(r^{4}+9r^{2}a^{2}+9a^{4} \right),
\ee
together with the gravitational energy flux given by
\be
{T}^{(0)(3)}=\frac{9L^{2}m^{2}a\Lambda(r^{2}+2a^{2})}{G_{0}r^{10}}.
\ee
The observer encounters the singularity at $r=0$, where the curvature components as well as the components of the gravitational 
stress-energy tensor all diverge.

\section*{5 Discussion}

An interesting result of this paper is the theoretical elucidation of the existence of a steady flux of gravitational
energy around a rotating mass.  It follows from equations (27) and (28) that
\be
{T}^{(0)(i)}/{T}^{(0)(0)} \sim 3 \frac{J}{M|{\bf x}|} ({\hat J}\times{\hat {\bf x}})^{i}
\ee
for large $|{\bf x}|$.  This means that the current of gravitational energy has speed $v_{g}=3a\sin\theta/r$ far from the
source, as can also be seen from the results of the last section.  In particular, for equatorial geodesics the ratio of equations (60) and (56) behaves as $3a/r$
when $r\gg a$.  On the other hand, the other results of the previous section, e.g. the diagonal form of equation (39), are not in
conflict with this conclusion since a detailed examination of the natural tetrad system (30)-(33) of the Kerr spacetime indicates
that the observer moves in the $\phi$-direction with speed $a/r$ far from the source, i.e.
\be
\frac{d\phi}{d\tau}=\frac{a}{(\Sigma\Delta)^{1/2}}
\ee
is the initial angular speed of the observer.  Moreover, the flow velocity $v_{g}{\bf \hat{\phi}}$ must vanish along the axis
of symmetry and this is consistent with the diagonal gravitational stress-energy tensor (48) for the test observer moving along the
$z$-axis.

In electrodynamics, the speed of the Poynting current is always less than or equal to the speed of light in
vacuum and the analogous result holds for the GEM tensor by equation (16). Therefore, it is interesting to investigate here how $v_{g}$ behaves as a function of $r$ in the equatorial plane of
the Kerr black hole as measured by the free ``radially'' infalling test observers.  As Figure 1 demonstrates,
\be
v_{g}=\frac{3a(r^{2}+a^{2})^{1/2}(r^{2}+2a^{2})}{r^{4}+6r^{2}a^{2}+6a^{4}}
\ee
monotonically increases with decreasing $r$ and reaches the speed of light at the $r=0$ singularity.

Finally, it is interesting to note that for a charged rotating mass --- such as a Kerr-Newman system --- there exist
both an electromagnetic Poynting flux and a gravitational flux of energy moving in essentially the same way about the rotating mass.
For instance, for the Kerr-Newman system we have that far from the source $v_{em} \sim 2a\sin\theta / r$ as $r \rightarrow \infty$.
The investigation of the physical consequences of the existence of  gravitational energy currents around rotating masses is beyond 
the scope of this work.

\appendix

\section*{Appendix A:  Bel-Robinson Tensor}

The gravitational superenergy tensor was first introduced independently by Bel and Robinson in close formal analogy
with Maxwell's stress-energy tensor \cite{1,3,13}.  In this regard, an important consideration was the invariance of
the Maxwell stress-energy tensor under duality rotations; therefore, a gravitational superenergy tensor was conceived that 
would be similarly invariant under duality rotations involving the gravitoelectric and gravitomagnetic fields.  Some aspects
of the early work in this direction can be found in \cite{14}. Further work on the Bel-Robinson tensor is contained in \cite{15}-\cite{17},
while other approaches to the stress-energy tensor for gravitation are discussed in \cite{18}.

Originally, spacetimes were considered with only a cosmological constant and in this case the tensor (12) already appears in the 
work of Bel \cite{3}.  Robinson proved that only in the Ricci-flat case would this tensor be totally symmetric and traceless.
Thus later work has been restricted to $R_{\mu\nu}=0$ and the Bel-Robinson tensor is thus defined by
\be
T_{\mu\nu\rho\sigma}=\frac{1}{2}\left( R_{\mu\xi\rho\zeta}R_{\nu\ \sigma}^{\ \xi\ \zeta} + R_{\mu\xi\sigma\zeta}R_{\nu\ \rho}^{\ \xi\ \zeta}
\right) - \frac{1}{16} g_{\mu\nu} g_{\rho\sigma} K.
\ee
That is, $\tilde{T}_{\mu\nu\rho\sigma} \rightarrow T_{\mu\nu\rho\sigma}$ since
\be
R_{\alpha\beta\rho\gamma}R^{\alpha\beta\ \gamma}_{\ \ \:\sigma}=\frac{1}{4} g_{\rho\sigma}K
\ee
for $R_{\mu\nu}=0$. Here $K$ is the Kretschmann scalar, i.e. $K =  R_{\mu\nu\rho\sigma}R^{\mu\nu\rho\sigma}$.

\section*{Appendix B:  Landau-Lifshitz Pseudotensor in Riemann Normal Coordinates}
  
The energy of a gravitational field --- if it can be defined at all --- is  nonlocal according to general relativity. On the
other hand,  the Bel-Robinson  tensor is locally defined. A connection could perhaps be established between  these concepts if
the energy-momentum pseudotensor of the gravitational  field is expressed in Riemann normal coordinates about a typical event
in  spacetime.

Let $x^{\mu}$ be the Riemann normal coordinates in the neighborhood of some  point (``origin") in spacetime; then,
\begin{eqnarray} g_{\mu\nu} = \eta_{\mu\nu} - {1 \over 3} R_{\mu\alpha\nu\beta}
\ x^{\alpha}  x^{\beta} + \cdots ,  \label{C1} 
\end{eqnarray}     
\begin{eqnarray}
\Gamma^{\mu}_{\nu\rho} =- {1 \over 3} (R^{\mu}_{\ \nu\rho\sigma} +
  R^{\mu}_{\ \rho\nu\sigma}) x^{\sigma} + \cdots .   \label{C2} 
\end{eqnarray}      The Landau-Lifshitz pseudotensor is quadratic in the connection coefficients by construction; therefore,
$t^{L-L}_{\mu\nu}$ is --- at the lowest order --- quadratic in Riemann normal coordinates. Hence,
\begin{eqnarray} t^{L-L}_{\mu\nu,\alpha\beta} =  {c^{4} \over 144 \pi G_0}
\Theta_{\mu\nu\alpha\beta} 
  + \cdots ,       \label{C3} 
\end{eqnarray}      where $\Theta_{\mu\nu\alpha\beta}$ is symmetric in its first and second  pairs of indices by construction
and is given by
\begin{eqnarray*}
\Theta_{\mu\nu\alpha\beta} =  {1 \over 2} (R^{\rho\sigma}_{\ \ \mu\alpha}  R_{\nu\sigma\rho\beta} +  R^{\rho\sigma}_{\ \
\mu\beta} R_{\nu\sigma\rho\alpha}) + {7 \over 2} (R_{\mu\rho\sigma\alpha} R_{\nu\ \ \beta}^{\ \rho\sigma} +
  R_{\mu\rho\sigma\beta} R_{\nu\ \ \alpha}^{\ \rho\sigma} )
\end{eqnarray*}     
\begin{eqnarray} -{3 \over 8} \eta_{\mu\nu} \eta_{\alpha\beta}  R_{\rho\sigma\kappa\delta} R^{\rho\sigma\kappa\delta}.
\label{C4} 
\end{eqnarray}      This expression should be compared and contrasted with equation (64)  that expresses the
Bel-Robinson tensor in a similar form.  There is no simple relationship between 
$\Theta_{\mu\nu\alpha\beta}$ and $T_{\mu\nu\alpha\beta}$;  however, one can show that
\begin{eqnarray}
\Theta_{\mu\nu\alpha\beta} -7 \ T_{\mu\nu\alpha\beta} = {1 \over 16} 
\eta_{\alpha\beta} \eta_{\mu\nu} K + {1 \over 4}  (R^{\rho\sigma}_{\ \
\mu\alpha}R_{\rho\sigma\nu\beta} + 
 R^{\rho\sigma}_{\ \ \mu\beta} R_{\rho\sigma\nu\alpha}) \label{C5} 
\end{eqnarray}      in Riemann normal coordinates.

\pagebreak[4]

\noindent
Figure 1 --- Plot of $v_g$ versus $r/a$ as given by equation (63), which represents the speed of the steady gravitational energy current in the equatorial plane of the Kerr system as perceived by observers falling freely from rest at spatial infinity.  It

 is important to note that $v_g \leq 1$ just as in electrodynamics.

\end{document}